\RequirePackage{fix-cm}
\documentclass{svjour3}                     
\smartqed  
\usepackage{graphicx}
\begin{document}

\title{The non-Abelian Weyl$-$Yang$-$Kaluza$-$Klein gravity model
}


\author{Halil Kuyrukcu
}


\institute{Physics Department,
B\"{u}lent Ecevit University,
67100, Zonguldak, Turkey\\
              \email{kuyrukcu@beun.edu.tr}}

\date{Received: date / Accepted: date}

\maketitle

\begin{abstract}
The Weyl$-$Yang gravitational gauge theory is investigated in the structure of a pure higher-dimensional non-Abelian Kaluza$-$Klein background. We construct the dimensionally reduced field equations and stress-energy-momentum tensors as well as the four dimensional modified  Weyl$-$Yang$+$Yang$-$Mills theory  from an arbitrary curved $internal$ space in the anholonomic frame which are the extensions of our previous model for the non-Abelian case. In particular, the coset space reduction is considered to explicitly obtain the interactions between the gravitational and the gauge fields. The resulting equations not only appear to be generalization of the well-established equations of non-Abelian theory but also contain  intrinsically the generalized gravitational source term which does not exist in the literature so far and the Yang$-$Mills force density which is exactly equivalent to the negative gradient of a Yang$-$Mills quadratic invariant.
\end{abstract}

\section{Introduction}

The Kaluza$-$Klein (KK) theories~\cite{Kaluza,Klein,Mandel,KleinN} base on the idea of the unification of well-known gravitational and $U(1)$ Abelian gauge interactions on a circle bundle with purely geometrical point of view in the four-dimensional ($4$D) effective theory, when extra space dimension is compactified (i.e. it is invisible through Kaluza's cylinder condition). With efforts of several authors~\cite{Einstein,Witt,Witt1,Rayski,Rayski1,Kerner,Trautman,Cho,Choo,Chooo,Chang}, this idea can naturally be generalized to the 	
situation of  non-Abelian gauge fields, which describe the strong and weak interactions in the particle physics. This approach leads to the most natural generalization of Einstein$-$Maxwell theory on a principal fibre bundle structure in the $(4+N)$-dimensional spacetime, i.e. the usual $4$D $external$ spacetime plus $N$-dimensional compact $internal$ sub-space which is usually preferred to be the coset space~\cite{Helgason,Kobayashi,Gilmore,witten77,forgacs,witten,kapetanakis} or the group manifold~\cite{scherk79} (and references therein). The gauge symmetry group of $external$ space comes remarkably from the isometry group of $internal$ space in the KK-type multi-dimensional unified field theories, and therefore they have been extensively discussed from many different angles over the years by other unification theories such as the supergravity~\cite{Duff} and superstring~\cite{Schwarz} theories (for a complete overview see~\cite{Appelquist,Bailin,Duff94,Overduin}).

To obtain equations of motions of Abelian or non-Abelian KK theories (most of gravitational theories), we conventionally prefer to employ well-established Hilbert$-$Einstein action which is linear in the curvature as a gravitational Lagrangian. However, there are  alternatively  Yang$-$Mills-style gravitational formulations (also known as gauge theories of gravitation, see Utiyama~\cite{utiyama}) which had already been considered very early by Weyl~\cite{weyl1} as a unified theory,  Lanczos~\cite{lanczos}, Lichnerowicz~\cite{lichnerowicz}, Stephenson~\cite{stephenson}, Higgs~\cite{higgs}, Kilmister and Newman~\cite{kilmister} for symmetric connection and metric, Loos~\cite{Loos}, Loos and Treat~\cite{Loos1} later improved by Yang~\cite{yang} for nonmetricity $GL(4,R)$ gauge theory, further investigated and discussed by Pavelle~\cite{pavelle}, Fairchild~\cite{fairchild,fairchild1,fairchild2}, more recently  Vassiliev~\cite{Vassiliev} as well as Mielke and Maggiolo~\cite{Maggiolo} in detail.

The Weyl$-$Yang gravitational Lagrangian is the quadratic in Riemann$-$Christofell tensor, and two distinct Einsteinian field equations are obtained by considering the metric tensor and affine connection that are independent dynamical field variables with ~\cite{baekler1,Szczyrba,mashhoon88,mashhoon91,Rose} and without torsion. The theory may be called Yang$-$Mills approach to gravity since the equations of motions are similarly achieved as those of the Yang$-$Mills~\cite{yangmills} or  the most well-known $U(1)$ classical electromagnetic theory. In fact, these Maxwellian field equations may appear to be a special limit of those given by Hehl and $et~al$~\cite{hehl78}. They have argued that nonvanishing torsion is also an independent variable within the framework of the theory so-called Poincar\'{e} Gauge theory of Gravity~\cite{hehl76} (references given there, and also see~\cite{daum,mielke2013}). Although the initial appeal for quadratic-type of alternative gravitational theory is faded as a gauge formulation of gravitational physics as well as the field equations of the theory that do not have an acceptable Newtonian limit together with the non-existence of a Birkhoff theorem~\cite{Blagojevic}, it is interesting in its own right and it is still prevailing as effective theories of modified gravity, specifically in quantum gravity~\cite{lausher2002} and loop quantum cosmology~\cite{cognola2013}. Recently, that simple quadratic gravitational Lagrangian model is shown to be very useful to overcome some cosmological problems in the real $4$D spacetime ~\cite{gerard,cook,gonzalez,chen,chen1,yeung,yeung1,yeung2,yeung3,yeung4,wang2013}. Moreover,  a great number of physical solutions of the theory along with unphysical ones are studied in the various combinations of the two sets of field equations by Pavelle~\cite{pavelle,pavelle1,pavelle2,pavelle3}, Thompson~\cite{thompson,thompson1}, Ni~\cite{ni,ni1}, Benn and $et~al$~\cite{benn81}, Mielke~\cite{mielke81}, Baekler and $et~al$~\cite{baekler1,baekler2}, Baekler~\cite{baekler}, Hsu and Yeung~\cite{Hsu}.  More recently,  physical plane-wave solutions of the theory are developed and discussed by Ba\c{s}kal~\cite{baskal} and Kuyrukcu~\cite{kuyrukcu} in  four and five dimensions, respectively.

In the present paper, we completely  generalize the methods and results of our previous work~\cite{sibelhalil} from $U(1)$ Abelian case to the non-Abelian case  without introducing any scalar boson fields by taking into account spinless and torsionless Weyl$-$Yang gravity model in the context of more than five dimensional pure KK theories. As it is well-known in KK theories, the $4$D matter is induced from geometry in higher dimensions that  the spacetime is empty, whereas in our approach, without loss of generality, the $4$D matter-spin source term  is induced from those that matter carrying energy-momentum  but not possessing any spin tensor. We also extend the reduced equations to more physical forms by taking  the compact $internal$ space as a homogeneous coset space which is necessary in KK theories. We can achieve this construction by following outline of our article. In section 2, we begin a brief review of Weyl$-$Yang gravity model in $(4+N)$ dimensions, to remind the reader some basic elements of that theory for convenient reading. In section 3, we perform a popular KK dimensional reduction procedure to obtain the modified $4$D  Weyl$-$Yang$+$Yang$-$Mills action associated with a pure $(4+N)$-dimensional Weyl$-$Yang gravitational Lagrangian without supplementary spin-matter fields which lead to  covariantly conservative energy-momentum tensor. In that respect, the dimensionally reduced field equations which naturally contain  the generalized gravitational source term that does not exist in the literature so far and the Yang$-$Mills force density which is exactly equivalent to the negative gradient of a Yang$-$Mills quadratic invariant and stress-energy tensors are simultaneously investigated in the anholonomic frame, and they are compared to our previous gravity model and standard higher-dimensional KK theories in section 4 and section 5, respectively. As a by-product, the coset space case of the theory is also included which leads to more physical results and some plain solutions of the field equations in detail. Finally, the last section demonstrates a brief discussion and conclusions obtained from our analysis. The appendix also contains a brief review to the major steps of non-Abelian KK framework in the anholonomic frame (as well as notational conventions) to introduce the reader some relations which we employ throughout this work.

\section{The Weyl$-$Yang gravity model in (4+N) dimensions}

The dynamics of Weyl$-$Yang theory of gravity is determined by the simple curvature-squared (quadratic in $\hat{R}_{ABCD}$) effective gravitational action (can be alternatively called Stephenson$-$Kilmister$-$Yang Lagrangian) on $M_{4+N}$ by following notations of Fairchild~\cite{fairchild}
\begin{eqnarray}\label{action}
\hat S_{WY}=-\frac{1}{4\hat \kappa^{2}}\int_{M_{4+N}}\hat{R}_{ABCD}^{2}\,\sqrt{-\hat g}\, d^{4}x\,d^{N}y+\hat {\mathcal{L}}_{m}[\hat\Gamma],
\end{eqnarray}
where the $\hat{\kappa}$ is coupling constant, and we use the shorthand notation $\hat{R}_{ABCD}^{2}\equiv\hat{R}_{ABCD}\hat{R}^{ABCD}$ with $\hat g=$det$(\hat g_{AB})$. A matter Lagrangian $\hat {\mathcal{L}}_{m}[\hat\Gamma]$ couples only with the gauge potential $\hat{\Gamma}^{A}\,_{BC}$ (i.e. the geometry) similar to the case of electromagnetic theory in the $(4+N)$-dimensional spacetime. This Riemannian model is introduced by Yang~\cite{yang} as an alternative approach to the Einstein's theory of General Relativity by employing perfect analogy with Yang$-$Mills theories. The Yang's theory is developed by Fairchild~\cite{fairchild} applying \textit{\`{a} la Palatini} variation (P-variation) principle~\cite{palatini,ferraris} to obtain Euler$-$Lagrange equations of the theory in a natural manner. The independent variation of the action (\ref{action}) with respect to the $(4+N)$-dimensional  arbitrary connection  $\hat{\Gamma}^{A}\,_{BC}$ which is chosen to be symmetric Christoffel connection,  and the symmetric metric $\hat{g}_{AB}$  gives the desired Yang$-$Mills-style field equations. Thus, from the connection variation, symbolically as ~$\delta \hat S_{\hat\Gamma}[\hat g,\hat\Gamma]=0$ with $\hat{\mathcal{L}}_{m}[\hat\Gamma]=0$, we obtain third-order nonlinear Yang's pure gravitational field equation
\begin{eqnarray}\label{fe}
\hat{D}_{A}\hat{R}^{A}\,_{BCD}=0.
\end{eqnarray}
This can also equivalently be written by using consecutive contractions and applying the Bianchi identities as
\begin{eqnarray}\label{fe2}
\hat{D}_{A}\hat{R}^{A}\,_{BCD}\equiv\hat{D}_{C}\hat{R}_{BD}-\hat{D}_{D}\hat{R}_{BC}=0.
\end{eqnarray}
Obviously, Yang's equation (\ref{fe}) is the generalized form of Einstein's field equations, and it contains naturally vacuum  Einsteinian solutions which are only $\hat{R}_{AB}=0$ and $\hat{R}_{AB}=\hat\lambda\hat g_{AB}$  for all constants $\hat\lambda$ as well as non-Einsteinian ones. Thus the equation (\ref{fe2}) is interpreted by Yang as the alternative form of the free-gravitational field equation.  The matter-source term of the test object $\hat{S}_{BCD}$~\cite{kilmister,fairchild,fairchild1,camenzind75,pavelle76,tseytlin}, antisymmetric in $C$ and $D$, may also be added to the right-hand side of the equation (\ref{fe}), namely
 \begin{eqnarray}\label{fes}
\hat{D}_{A}\hat{R}^{A}\,_{BCD}=4\hat\kappa^{2}\hat{S}_{BCD}\equiv \delta \hat {\mathcal{L}}^{m}_{\hat\Gamma}[\hat\Gamma].
\end{eqnarray}
This gravitational current term yields the cyclic symmetry and  covariant conservation properties~\cite{oktem} as expected from a source term
\begin{equation}\label{identities}
\hat S_{[BCD]}=0, \qquad\qquad\qquad \hat D_{B}\hat S^{B}\,_{CD}=0,
\end{equation}
and it is interpreted as the covariant derivative of the Einstein's matter stress-energy tensor by Kilmister~\cite{kilmister66} later Camenzind~\cite{camenzind75} without a variational principle and more conveniently  as the canonical spin-density of the matter fields by Fairchild~\cite{fairchild} and very early by Loos~\cite{Loos}. We can also rewrite sourceless field equations (\ref{fe}) in many other ways, see~\cite{baekler,garecki}.

From the metric variation~$\delta \hat S_{\hat g}[\hat g,\hat\Gamma]\equiv(\hat\kappa^{2}/2)\hat{T}_{AB}$, we arrive at the symmetric second-rank gravitational matter-energy-momentum tensor in the following form
\begin{equation}\label{set}
\hat{T}_{AB}\equiv\hat{R}_{ACDE}\hat{R}_{B}\,^{CDE}
-\frac{1}{4}\hat{g}_{AB}\hat{R}^{2}_{CDEF},
\end{equation}
which is called Stephenson equation for the case $\hat{T}_{AB}=0$~\cite{stephenson}. If there are not only spacetime torsion but also canonical source term,  the equation (\ref{set}) automatically satisfies a conservation law (divergence-free) of the form $\hat D_{A}\hat{T}^{A}\,_{B}=0$  but not traceless $\hat{T}\equiv\hat{T}_{A}\,^{A}\neq0$ in higher dimensions. It is actually $\hat{T}=-(N/4)\hat{R}^{2}_{ABCD}$  and totally traceless only $N=0$ case, i.e. for the usual $4$D spacetime. The $\hat{T}_{AB}$ is also rewritten
\begin{equation}\label{con}
\hat{T}_{AB}=2\hat C_{ACBD}\hat R^{CD}+\frac{1}{3}\hat{R}(\hat{R}_{AB}-\frac{1}{4}\hat g_{AB}\hat R),
\end{equation}
in terms of the traceless Weyl conformal tensor $\hat C_{ACBD}$~\cite{fairchild,baekler,oktem}. If $\hat{T}_{AB}=0$ one can prove that two empty-space solutions above also satisfy equation (\ref{con}) but we will not discuss this interesting case for our model. The equation (\ref{set}) is indeed the contraction form of the well-known Bel$-$Robinson superenergy tensor~\cite{bel,bel1,robinson,robinson1,mashhoon}, and it appears perfectly analogous to stress-energy of classical electromagnetic theories as well.

\section{The reduction of the Stephenson$-$Kilmister$-$Yang  curvature}

The method of dimensional reduction, from higher-dimensional theory to the real $4$D spacetime, 	
brings the types of gravitational and gauge fields together with the forms of interactions between the constituent fields into the clear. The reduced form of the $(4+N)$-dimensional quadratic curvature after performing dimensional reduction procedure is given by
\begin{eqnarray}\label{expr2}
 \hat{R}_{ABCD}^{2}
= \hat{R}_{\mu\nu\lambda\sigma}^{2}
+4(\hat{R}_{i\nu\lambda\sigma}^{2}
+\hat{R}_{i\nu k\sigma}^{2}
+\hat{R}_{ijk\sigma}^{2})
+2\hat{R}_{ij\lambda\sigma}^{2}
+\hat{R}_{ijkl}^{2}.
\end{eqnarray}
The substitutions of Riemann's components (in equation (\ref{riem})) into equation (\ref{expr2}) with a tedious computations and help of the relation
\begin{eqnarray}
R_{\mu\nu\lambda\sigma}F^{\alpha \mu\lambda}F^{\beta \nu\sigma}=\frac{1}{2}R_{\mu\nu\lambda\sigma}F^{\alpha \mu\nu}F^{\beta \lambda\sigma},
\end{eqnarray}
gives the dimensionally reduced invariant in the following form
\begin{eqnarray}\label{rlag}
\hat{R}_{ABCD}^{2}
&=&R_{\mu\nu\lambda\rho}^{2}-\frac{3}{2}\xi^{\alpha i}\xi^{\beta}_{i} R_{\mu\nu\lambda\rho}F^{\alpha\mu\nu}F^{\beta\lambda\rho}
+\xi^{\alpha i}\xi^{\beta}_{i}\mathcal{D}_{\mu}F^{\alpha}_{\nu\lambda}
\mathcal{D}^{\mu}F^{\beta\nu\lambda}\nonumber\\&&
+\frac{1}{8}\xi^{\alpha i}\xi^{\beta}_{i}
\xi^{\gamma j}\xi^{\eta}_{j}
(
\mathcal{F}^{4}_{\alpha\gamma\beta\eta}
+3\mathcal{F}^{2}_{\alpha\gamma}\mathcal{F}^{2}_{\beta\eta}
+4\mathcal{F}^{4}_{\alpha\gamma\eta\beta})
\\&&
+3(D_{i}\xi^{\alpha}_{j}D^{i}\xi^{\beta j}\mathcal{F}^{2}_{\alpha\beta}+\xi^{\alpha}_{i}\xi^{\beta}_{j}D^{i}\xi^{\gamma j}\mathcal{F}^{3}_{\alpha\beta\gamma})
+R_{ijkl}^{2},\nonumber
\end{eqnarray}
where the following abbreviations are introduced
\begin{eqnarray}
&&\mathcal{F}_{\alpha\beta}^{2}=F^{\alpha}_{\mu\nu}F^{\beta\mu\nu},\qquad\qquad\quad
\mathcal{F}_{\alpha\beta\gamma}^{3}=F^{\alpha}_{\mu\lambda}F^{\beta\lambda\sigma}F^{\gamma}_{\sigma}\,^{\mu},
\nonumber\\
&&\mathcal{F}^{3~\alpha\beta\gamma}_{\mu\nu}=F^{\alpha}_{\mu\lambda}F^{\beta\lambda\sigma}F^{\gamma}_{\sigma\nu},\qquad\qquad\quad
\mathcal{F}_{\alpha\beta\gamma\eta}^{4}=F^{\alpha}_{\mu\lambda}F^{\beta\lambda\sigma}F^{\gamma}_{\sigma\nu}
F^{\eta\nu\mu},
\\ &&\mathcal{F}^{4~\alpha\beta\gamma\eta}_{\mu\nu}=F^{\alpha}_{\mu\lambda}F^{\beta\lambda\sigma}F^{\gamma}_{\sigma\rho}
F^{\eta\rho}\,_{\nu}.\nonumber
\end{eqnarray}
Furthermore, in equation (\ref{rlag}) and what follows, the gauge $\mathcal{D}_{[\mu}F^{\alpha}_{\nu\lambda]}=0$ and the gravitational $D_{[\tau}R_{\mu\nu]\lambda\sigma}=0$ Bianchi identities are used, whenever we need them. We also recognize the fact that, the equation (\ref{rlag}) is a common result for all the basis, because $\hat{R}_{ABCD}^{2}$ is clearly an invariant. We can overcome the term $D_{i}\xi^{\alpha}_{j}$ by taking Wu and Zee's assumption \cite{Yong-shi} into account, then we get a simple formula for covariant derivative of Killing vector fields as
\begin{equation}\label{kil2}
D_{i}\xi^{\alpha}_{j}=\frac{1}{2}f^{\alpha\beta\gamma}\xi^{\beta}_{i}\xi^{\gamma}_{j}.
\end{equation}
However, in this case the $internal$ manifold is reduced to be homogeneous space which is first discussed by Luciani \cite{Luciani} in KK theories. Thus, the extra space can now be written as the symmetric coset space $M_{N}=G/H$. $H$ is defined to be the isotropy subgroup of $G$, $H\subset G$ with $N=dim(G)-dim(H)$. All the Killing vector terms in the equation (\ref{rlag}) are then disappeared, just as expected, by employing equations (\ref{kil2}) and (\ref{kill}) with $c=1$
\begin{eqnarray}\label{rlag1}
\hat{R}_{ABCD}^{2}
&=&R_{\mu\nu\lambda\rho}^{2}-\frac{3}{2} R_{\mu\nu\lambda\rho}F^{\alpha\mu\nu}F^{\alpha\lambda\rho}
+\frac{1}{8}[\mathcal{F}^{4}_{\alpha\beta\alpha\beta}
+3(\mathcal{F}^{2}_{\alpha\beta})^{2}
+4\mathcal{F}^{4}_{\alpha\beta\beta\alpha}]
\nonumber\\
&&
+\frac{3}{4}(f^{\alpha\gamma\eta}f^{\beta\gamma\eta}\mathcal{F}^{2}_{\alpha\beta}
+2f^{\alpha\beta\gamma}\mathcal{F}^{3}_{\alpha\beta\gamma})
+\mathcal{D}_{\mu}F^{\alpha}_{\nu\lambda}\mathcal{D}^{\mu}F^{\alpha\nu\lambda}
+R_{ijkl}^{2}.
\end{eqnarray}
It is easy to see that, in addition to the standard $4$D Weyl$-$Yang Lagrangian term $R_{\mu\nu\lambda\rho}^{2}$, the effective $4$D action (\ref{rlag1}) also contains the well-known $RF^{2}$-type term~\cite{liu} (and references therein) which describes a non-minimal coupling between curvature and the non-Abelian gauge field and the self-interacting Yang$-$Mills field invariants which are the cubic term $F^{3}$ and the quartic terms $F^{4}$ as well. The invariant form $F^{3}$ appears only in non-Abelian theories, and it is first studied by Alekseev and Arbuzov~\cite{Alekseev82,Alekseev84}. The ordinary Yang$-$Mills Lagrangian term $ \mathcal{F}^{2}_{\alpha\alpha}$ can be obtained from the $f^{\alpha\gamma\eta}f^{\beta\gamma\eta}\mathcal{F}^{2}_{\alpha\beta}$ term, if we normalize the structure constants such that $f^{\alpha\gamma\eta}f^{\beta\gamma\eta}=(N-1)\delta^{\alpha\beta}$. Then for the Abelian case $N=1$ this term goes trivially to zero. By virtue of Leibniz rule, the term with covariant derivative of gauge field tensor may be rewritten symbolically as $(\mathcal{D}F)^{2}=\mathcal{D}(F\mathcal{D}F)-F\mathcal{D}^{2}F$ of which first term is  total derivative so it can be dropped from the action, thus we just have the $F\mathcal{D}^{2}F$-type interactions. Finally, the last term $R_{ijkl}^{2}$ can be interpreted as the cosmological constant term as well as the $4$D Ricci scalar of $internal$ space $R(y)$ (see appendix) or it can be just ignored  so that it does not contain any physical fields~\cite{Huang}. The invariant $R_{ijkl}^{2}$ is actually determined by the dimension number of $internal$  space $N$ as $R_{ijkl}^{2}=(1/8)N(N-1)$ for coset space. Then the right-hand side of equation (\ref{rlag1}) depends only on $external$ coordinates just like the equation (\ref{scalar}).  As a conclusion, the
equation (\ref{rlag1}) leads to a modified $4$D Weyl$-$Yang$+$Yang$-$Mills action with higher-order terms, and in fact it turns naturally out to be a part of second-order Euler$-$Poincar\'{e} (Gauss$-$Bonnet) curvature invariant~\cite{Huang,müller,dereli,kimm}. We should also emphasise that, although the dimensional reduction of quadratic Lagrangian from $(4+N)$ to four dimensions is discussed very early by Cho and $et~al$~\cite{Choo}, they have not obtained proper Yang$-$Mills term $\mathcal{F}^{2}_{\alpha\alpha}$  but rather discussed only third-order $(\mathcal{D}F)^{2}$ term.

\section{The reduction of the source-free Yang's equation}

Now, we are ready to work out the reduced components of source-free field equations $\hat{D}_{A}\hat{R}^{A}\,_{BCD}=0$ in (\ref{fe}) by taking into account the KK  reduction scheme, as a traditional way. Thus, in the $(4+N)$ decomposition we have the six different equations of motions (one $4$D part and five $N$-dimensional parts) which are complementary with each other
\begin{eqnarray}
&&\hat{D}_{A}\hat{R}^{A}\,_{\nu\lambda\sigma}=0,\label{fee1} \\
&&\hat{D}_{A}\hat{R}^{A}\,_{i\lambda\sigma}=0,\qquad\qquad
\hat{D}_{A}\hat{R}^{A}\,_{\nu j\sigma}=0,\label{fee2}\\
&&\hat{D}_{A}\hat{R}^{A}\,_{i\lambda k}=0,\qquad\qquad
\hat{D}_{A}\hat{R}^{A}\,_{\nu j k}=0,\label{fee3}\\
&&\hat{D}_{A}\hat{R}^{A}\,_{ijk}=0.\label{fee4}
\end{eqnarray}
After some lengthy but careful manipulations, on account of main equation (\ref{riem}) together with the Killing vector relations (equations (\ref{ke1}) and (\ref{ke2})) and the dual basis equation (\ref{dual}), one obtains the $4$D part of the $(4+N)$-dimensional Yang's equation in (\ref{fee1}) as the $4$D Yang's equation with the current term
\begin{eqnarray}\label{rfe1}
D_{\mu}R^{\mu}\,_{\nu\lambda\sigma}=S_{\nu\lambda\sigma},
\end{eqnarray}
where
\begin{eqnarray}\label{source}
 S_{\nu\lambda\sigma}(x,y)&=&
\frac{1}{2}\xi^{\alpha i}\xi^{\beta}_{i}[
 J^{\alpha}_{\nu}F^{\beta}_{\lambda\sigma}
+\frac{1}{2}(J^{\alpha}_{\lambda}F^{\beta}_{\nu\sigma}
-J^{\alpha}_{\sigma}F^{\beta}_{\nu\lambda})
 +\mathcal{D}_{\sigma}
(F^{\alpha}_{\lambda\mu}F^{\beta\mu}\,_{\nu})\nonumber\\
& &
-\mathcal{D}_{\lambda}
(F^{\alpha}_{\sigma\mu}F^{\beta\mu}\,_{\nu})].
\end{eqnarray}
Here, $J^{\alpha}_{\nu}$ is usual non-zero four-current term of Yang$-$Mills theory, i.e. $J^{\alpha}_{\nu}=\mathcal{D}_{\mu}F^{\alpha\mu}\,_{\nu}$. The term $S_{\nu\lambda\sigma}(x,y)$ may be interpreted as the gravitational source like-term (gauge current term) of the $4$D matter, and it consists of various combinations of $F$ and $\mathcal{D}F$ terms. It is not difficult to conjecture that, the equation (\ref{source}) automatically satisfies the identities (\ref{identities}). This result (\ref{rfe1}) is very valuable from the physical point of view in case the $4$D matter-spin tensor term is induced from  higher dimensions that matter carrying energy-momentum  but not possessing any spin tensor. In this sense, our approach is formally similar to the KK theories.  Besides, we can remark that the first reduced field equation (\ref{rfe1}) governs mainly the  gravitational fields, since the $DR$ term have the highest order derivative of the $external$ metric $g_{\mu\nu}(x)$.

The reduced form of field equations (\ref{fee2}), however, give more complicated relations
\begin{eqnarray}\label{rfe2}
\xi^{\alpha}_{i}\mathcal{D}_{\mu}\mathcal{D}^{\mu}F^{\alpha}_{\lambda\sigma}&&=
-\xi^{\alpha}_{i}R_{\mu\nu\lambda\sigma}F^{\alpha\mu\nu}
+2\xi^{\alpha j}D_{j}\xi^{\beta}_{i} (F^{\alpha}_{\lambda\mu}F^{\beta\mu}\,_{\sigma}
-F^{\beta}_{\lambda\mu}F^{\alpha\mu}\,_{\sigma})\nonumber\\&&
-2F^{\alpha}_{\lambda\sigma}D_{j}D^{j}\xi^{\alpha}_{i}
+\frac{1}{2}\xi^{\alpha}_{i}\xi^{\beta j}\xi^{\gamma}_{j}(\mathcal{F}^{2}_{\alpha\beta}F^{\gamma}_{\lambda\sigma}
+\mathcal{F}^{3~\alpha\beta\gamma}_{\sigma\lambda}
-\mathcal{F}^{3~\alpha\beta\gamma}_{\lambda\sigma}
),
\end{eqnarray}
\begin{eqnarray}\label{rfe3}
\xi^{\alpha}_{i} \mathcal{D}_{\mu}\mathcal{D}_{\sigma}F^{\alpha\mu}\,_{\nu}&=&
-\xi^{\alpha}_{i}R_{\mu\nu\tau\sigma}F^{\alpha\tau\mu}
 +2\xi^{\alpha j}D_{j}\xi^{\beta}_{i}F^{\alpha}_{\sigma\mu}F^{\beta\mu}\,_{\nu}
 +F^{\alpha}_{\nu\sigma}D_{j}D^{j}\xi^{\alpha}_{i}
 \nonumber\\&&
-\frac{1}{4}\xi^{\alpha}_{i}\xi^{\beta j}\xi^{\gamma}_{j}(\mathcal{F}^{2}_{\alpha\beta}F^{\gamma}_{\nu\sigma}
-2\mathcal{F}^{3~\alpha\beta\gamma}_{\nu\sigma}).
\end{eqnarray}
The equations (\ref{rfe2}) and (\ref{rfe3}) basically govern the Yang$-$Mills gauge fields in view of $\mathcal{D}\mathcal{D}F$, and they include non-minimal direct $RF$-type couplings together with the cubic terms $F^{3}$. Additionally,  we can obtain the covariant derivative of the non-Abelian currents $\mathcal{D}_{\mu}J^{\alpha}_{\nu}$ by summing  two equations ((\ref{rfe2}) and (\ref{rfe3})) above as well as employing  Bianchi identities with the following relation
\begin{eqnarray}\label{rel}
\xi^{\alpha}_{i}[\mathcal{D}_{\mu},\mathcal{D}_{\nu}]F^{\alpha\lambda}\,_{\sigma}
=\xi^{\alpha}_{i}(R^{\lambda}\,_{\tau\mu\nu}F^{\alpha \tau}\,_{\sigma}-R^{\tau}\,_{\sigma \mu\nu}F^{\alpha\lambda}\,_{\tau})
+2\xi^{\beta j}D_{j} \xi^{\alpha}_{i} F^{\alpha}_{\mu\nu}F^{\beta\lambda}\,_{\sigma}.
\end{eqnarray}
Hence, the result is
\begin{eqnarray}
\xi^{\alpha}_{i} \mathcal{D}_{\mu}J^{\alpha}_{\nu}=
\frac{1}{4}\xi^{\alpha}_{i}(\mathcal{F}^{2}_{\alpha\beta}\mathcal{F}^{\beta}_{\mu\nu}-\mathcal{F}^{3~\alpha\beta\beta}_{\mu\nu}
+2\mathcal{F}^{3~\alpha\beta\beta}_{\nu\mu})
+\xi^{\alpha}_{i}F^{\alpha}_{\nu\lambda}R^{\lambda}\,_{\mu}
+\xi^{\alpha}_{j}F^{\alpha}_{\mu\nu}R_{i}\,^{j}.
\end{eqnarray}
It is easy to prove that the contraction of free indices gives precisely  the covariant conservation law for $external$ current such as $\mathcal{D}_{\mu}J^{\alpha\mu}=0$.

In an equivalent way, the field equations with two $internal$ space indices (\ref{fee3}) are respectively reduced as
 \begin{eqnarray}
&&\xi^{\alpha}_{i}\xi^{\beta}_{k}(\mathcal{D}_{\lambda}\mathcal{F}^{2}_{\alpha\beta}
+F^{\alpha}_{\lambda\tau}J^{\beta\tau})
+2J^{\alpha}_{\lambda}D_{k}\xi^{\alpha}_{i}=0,\label{rfe4}\\&&
\xi^{\alpha}_{i}\xi^{\beta}_{k}(F^{\alpha}_{\lambda\tau}J^{\beta\tau}
-F^{\beta}_{\lambda\tau}J^{\alpha\tau})
+4J^{\alpha}_{\lambda}D_{k}\xi^{\alpha}_{i}=0 .\label{rfe5}
\end{eqnarray}
Another interesting result is obtained, if we eliminate the common term $J^{\alpha}_{\lambda}D_{k}\xi^{\alpha}_{i}$ of equations (\ref{rfe4}) and (\ref{rfe5}) by adding the equations together. Then, we read
\begin{eqnarray}
\xi^{\alpha}_{i}\xi^{\beta}_{k}[2\mathcal{D}_{\lambda}\mathcal{F}^{2}_{\alpha\beta}
+F^{\alpha}_{\lambda\tau}J^{\beta\tau}
+F^{\beta}_{\lambda\tau}J^{\alpha\tau}]=0,
\end{eqnarray}
which alternatively corresponds to
\begin{eqnarray}\label{lfd}
f_{\lambda}^{\alpha\beta}+f_{\lambda}^{\beta\alpha}
=-2\mathcal{D}_{\lambda}\mathcal{F}^{2}_{\alpha\beta}.
\end{eqnarray}
Here, $f_{\lambda}^{\alpha\alpha}=F^{\alpha}_{\lambda\tau}J^{\alpha\tau}$ is confidentially interpreted as a Lorentz-like force density of the non-Abelian theory. Although such  terms generally emerge from the higher-dimensional geodesic equations~\cite{Kerner,Orzalesi}, they naturally appear in the reduced field equations of our approach. It is difficult to understand physical meaning of this circumstance (\ref{lfd}), however,  we can at least theoretically say that the Yang$-$Mills invariant $\mathcal{F}^{2}_{\alpha\beta}$ is defined as a field whose gradient is equal and opposite to the generalized density of Lorentz force which is conservative. It is worthwhile to mention that in the language of fluid mechanics, the $f$ corresponds to a force density and the invariant $\mathcal{F}^{2}$ to a pressure term~\cite{Landau}.

Finally, from the last field equation (\ref{fee4}), we obtain the Yang's equation with source term again, but in this case it is for the $internal$ space
\begin{equation}\label{rfe6}
D_{l}R^{l}\,_{ijk}=S_{ijk},
\end{equation}
where the unphysical source-like term is found to be
\begin{equation}\label{source1}
S_{ijk}(x,y)=-\frac{1}{4}\mathcal{F}^{2}_{\alpha\beta}(
\xi^{\alpha}_{k}D_{j}\xi^{\beta}_{i}
-\xi^{\alpha}_{j}D_{k}\xi^{\beta}_{i}
-2\xi^{\alpha}_{i}D_{k}\xi^{\beta}_{j}),
\end{equation}
which provides the same  conservation and symmetry conditions as the term $S_{\nu\lambda\sigma}$ in (\ref{source}). In the absence of Yang$-$Mills fields we only have source-free Yang's equations which correspond to the pure gravitational interactions for $external$ $D_{\mu}R^{\mu}\,_{\nu\lambda\sigma}=0$ and $internal$ $D_{l}R^{l}\,_{ijk}=0$ space, respectively.

Attention will now be turned to investigate the couplings between the components of $(4+N)$-dimensional Ricci tensor $\mathcal{P}_{\mu\nu}$, $\mathcal{Q}_{i\nu}$, $\mathcal{U}_{ij}$ in equations (\ref{skk1})$-$(\ref{skk3}) and the $2$-forms gauge fields $F$. By virtue of alternative field equation (\ref{fe2}), one can recognize the fact that $\{\mathcal{P}_{\mu\nu}, \mathcal{Q}_{i\nu}, \mathcal{U}_{ij}\}$-set is embedded in the Weyl$-$Yang field equations of non-Abelian theory  (\ref{fee1})$-$(\ref{fee4}). Hence, from the reduced equations ((\ref{rfe1}) with (\ref{source}), (\ref{rfe2}), (\ref{rfe3}), (\ref{rfe4}), (\ref{rfe5}) and (\ref{rfe6}) with (\ref{source1})) we deduce the following compact embedded equations
\begin{eqnarray}
&&\mathcal{D}_{\lambda}\mathcal{P}_{\nu\sigma}
-\mathcal{D}_{\sigma}\mathcal{P}_{\nu\lambda}
+\frac{1}{4}\xi^{\alpha i}(F^{\alpha}_{\nu\lambda}\mathcal{Q}_{i\sigma}
-F^{\alpha}_{\nu\sigma}\mathcal{Q}_{i\mu\lambda}
-2F^{\alpha}_{\lambda\sigma}\mathcal{Q}_{i\nu})=0,\label{emb1}\\
&&\mathcal{D}_{\lambda}\mathcal{Q}_{i\sigma}
-\mathcal{D}_{\sigma}\mathcal{Q}_{i\lambda}
+\xi^{\alpha}_{i}(F^{\alpha\mu}\,_{\sigma}\mathcal{P}_{\mu\lambda}
-F^{\alpha\mu}\,_{\lambda}\mathcal{P}_{\mu\sigma})
+2\xi^{\alpha j}F^{\alpha}_{\lambda\sigma}\mathcal{U}_{ij}=0,\label{cfe1}\\
&&\mathcal{D}_{\sigma}\mathcal{Q}_{i\nu}
+\xi^{\alpha}_{i}F^{\alpha\mu}\,_{\nu}\mathcal{P}_{\mu\sigma}
-\xi^{\alpha j}F^{\alpha}_{\nu\sigma}\mathcal{U}_{ij}=0,\label{cfe2}\\
&&\mathcal{D}_{\lambda}\mathcal{U}_{ik}
+\frac{1}{4}(\xi^{\alpha}_{i} F^{\alpha}_{\lambda}\,^{\tau}\mathcal{Q}_{k\tau}
+2D_{k}\mathcal{Q}_{i\lambda})
= 0,\label{emb3}\\
&&
D_{k}\mathcal{Q}_{j\nu}
+\frac{1}{4}F^{\alpha}_{\nu}\,^{\tau}(\xi^{\alpha}_{j} \mathcal{Q}_{k\tau}
-\xi^{\alpha}_{k} \mathcal{Q}_{j\tau})= 0 ,\label{emb4}\\
&&
D_{i}\mathcal{U}_{kj}-D_{j}\mathcal{U}_{ki}= 0 \label{emb2}.
\end{eqnarray}
Here, we made use of identities as mentioned before and the relation $D_{j}D_{k}\xi^{\alpha}_{l}=\xi^{\alpha }_{i}R^{i}\,_{jkl}$~\cite{weinberg}. The equation (\ref{cfe1}) can be obtained from the equation (\ref{cfe2}), thus it is not essential. The equations (\ref{emb1})$-$(\ref{emb2}) contain the non-Abelian covariant derivative of $\{\mathcal{P}_{\mu\nu}, \mathcal{Q}_{i\nu}, \mathcal{U}_{ij}\}$-set, as well as the ordinary derivative of $\{\mathcal{Q}_{i\nu}, \mathcal{U}_{ij}\}$-set and various $\mathcal{P}F$, $\mathcal{Q}F$-type coupling terms. Generalized field equations above are welcome from another point of view that any solution to the $\{\mathcal{P}_{\mu\nu}, \mathcal{Q}_{i\nu}, \mathcal{U}_{ij}\}$-set solves the embedded equations (\ref{emb1})$-$(\ref{emb2}) in a natural manner.

However, there is a mismatch between the left-hand side of the field equation (\ref{rfe1}) which depends only on $external$ coordinates $x$ and the right-hand side which depends both on $external$  and $internal$  coordinates $x,y$ just as $\hat R(x,y)$ in equation (\ref{scalar}). Therefore, not only to avoid this problem but also to reduce equations (\ref{emb1})$-$(\ref{emb2}) to the simpler and physical forms, we restrict our considerations such that the $internal$ space is taken to be the homogeneous coset space the type $G/H$. It is easy to see that the source term (\ref{source}) with equation (\ref{kill}) and $c=1$ yields
\begin{eqnarray}\label{n0}
 S_{\nu\lambda\sigma}(x)=\frac{1}{2}[
 J^{\alpha}_{\nu}F^{\alpha}_{\lambda\sigma}
+\frac{1}{2}(J^{\alpha}_{\lambda}F^{\alpha}_{\nu\sigma}
-J^{\alpha}_{\sigma}F^{\alpha}_{\nu\lambda})
 +\mathcal{D}_{\sigma}
(F^{\alpha}_{\lambda\mu}F^{\alpha\mu}\,_{\nu})
-\mathcal{D}_{\lambda}
(F^{\alpha}_{\sigma\mu}F^{\alpha\mu}\,_{\nu})].\nonumber\\
\end{eqnarray}
Then both sides of the equation (\ref{n0}) are only associated  with $external$ $x$ coordinates as it should be. Although the same 	
difficulty occurs in the unphysical source-like term (\ref{source1}), this is not a problem after the coset space reduction procedure.

Next by using equation (\ref{kil2}) and  reduced form of equation (\ref{rel})
\begin{eqnarray}
[\mathcal{D}_{\mu},\mathcal{D}_{\nu}]F^{\alpha\lambda}\,_{\sigma}
=R^{\lambda}\,_{\tau\mu\nu}F^{\alpha \tau}\,_{\sigma}-R^{\tau}\,_{\sigma \mu\nu}F^{\alpha\lambda}\,_{\tau}
+f^{\beta\gamma\alpha}F^{\beta}_{\mu\nu}F^{\gamma\lambda}\,_{\sigma}.
\end{eqnarray}
the first embedded equation (\ref{emb1}) becomes
\begin{eqnarray}\label{n1}
&&\mathcal{D}_{\lambda}
(R_{\sigma\nu}-\frac{1}{2}F^{\alpha}_{\sigma\mu}F^{\alpha}_{\nu}\,^{\mu})
-\mathcal{D}_{\sigma}
(R_{\lambda\nu}-\frac{1}{2}F^{\alpha}_{\lambda\mu}F^{\alpha}_{\nu}\,^{\mu})
\nonumber\\& &
+\frac{1}{4}[F^{\alpha}_{\nu\lambda}\mathcal{D}_{\mu}F^{\alpha\mu}\,_{\sigma}
-F^{\alpha}_{\nu\sigma}\mathcal{D}_{\mu}F^{\alpha\mu}\,_{\lambda}
-2F^{\alpha}_{\lambda\sigma}\mathcal{D}_{\mu}F^{\alpha\mu}\,_{\nu}]=0.
\end{eqnarray}
The second (\ref{cfe1}) and the third (\ref{cfe2}) equations simultaneously yield
\begin{eqnarray}\label{n22}
&&\xi^{\alpha}_{i}[\mathcal{D}_{\lambda}(\mathcal{D}_{\mu}F^{\alpha\mu}\,_{\sigma})
-\mathcal{D}_{\sigma}(\mathcal{D}_{\mu}F^{\alpha\mu}\,_{\lambda})
+F^{\alpha\mu}\,_{\sigma}( R_{\mu\lambda}-\frac{1}{2}F^{\beta}_{\mu\tau}F^{\beta}_{\lambda}\,^{\tau})
\nonumber\\&&
-F^{\alpha\mu}\,_{\lambda}( R_{\mu\sigma}-\frac{1}{2}F^{\beta}_{\mu\tau}F^{\beta}_{\sigma}\,^{\tau})
+\frac{1}{2}F^{\beta}_{\lambda\sigma}(\mathcal{F}^{2}_{\alpha\beta}
-f^{\eta\gamma\alpha}f^{\beta\gamma\eta})]= 0,
\end{eqnarray}
\begin{equation}\label{n3}
\xi^{\alpha}_{i}[\mathcal{D}_{\sigma}(\mathcal{D}_{\mu}F^{\alpha\mu}\,_{\nu})
 +F^{\alpha\mu}\,_{\nu}( R_{\mu\sigma}-\frac{1}{2}F^{\beta}_{\mu\tau}F^{\beta}_{\sigma}\,^{\tau})
+\frac{1}{4}F^{\beta}_{\nu\sigma}(\mathcal{F}^{2}_{\alpha\beta}
-f^{\eta\gamma\alpha}f^{\beta\gamma\eta})]= 0,
\end{equation}
and the equations (\ref{emb3}) and (\ref{emb4}) similarly imply that
\begin{eqnarray}\label{n4}
\xi^{\alpha}_{i}\xi^{\beta}_{k}[
\mathcal{D}_{\lambda}(\mathcal{F}^{2}_{\alpha\beta}
-f^{\eta\gamma\alpha}f^{\beta\gamma\eta})
+F^{\alpha}_{\lambda\tau}\mathcal{D}_{\mu}F^{\beta\mu\tau}
+f^{\gamma\beta\alpha}\mathcal{D}_{\mu}F^{\gamma\mu}\,_{\lambda}]= 0,
\end{eqnarray}
\begin{eqnarray}\label{n5}
\xi^{\alpha}_{j}\xi^{\beta}_{k}[F^{\alpha}_{\nu\tau}\mathcal{D}_{\mu}F^{\beta\mu\tau}
-F^{\beta}_{\nu\tau}\mathcal{D}_{\mu}F^{\alpha\mu\tau}
+2f^{\gamma\beta\alpha}\mathcal{D}_{\mu}F^{\gamma\mu}\,_{\nu}]= 0.
\end{eqnarray}
Finally, the last equation (\ref{emb2}), on the other hand, can be expressed in the following form
\begin{eqnarray}\label{n6}
&&\xi_{i}^{\alpha}\xi_{j}^{\beta}\xi_{k}^{\gamma}[f_{\eta\alpha\gamma}(\mathcal{F}^{2}_{\beta\eta}
-f^{\zeta\pi\beta}f^{\eta\pi\zeta})-f_{\eta\beta\gamma}(\mathcal{F}^{2}_{\alpha\eta}
-f^{\zeta\pi\alpha}f^{\eta\pi\zeta})\nonumber\\&&
+2f_{\eta\alpha\beta}(\mathcal{F}^{2}_{\eta\gamma}
-f^{\zeta\pi\eta}f^{\gamma\pi\zeta})]=0,
\end{eqnarray}
where the expression $R_{ij}=-(1/4)f^{\eta\gamma\alpha}f^{\beta\gamma\eta}\xi_{i}^{\alpha}\xi_{j}^{\beta}$ is also used \cite{Yong-shi}.

If we remove all the Killing vector terms from embedded equations (\ref{n22})$-$(\ref{n6}), then all equations are independent of the $internal$ coordinates. These seem to be physically attractive, because it is possible to easily recognize the following plain solutions in the reduced equations
\begin{eqnarray}\label{n19}
R_{\mu\nu}=\frac{1}{2}F^{\alpha}_{\mu\tau}F^{\alpha}_{\nu}\,^{\tau},
\qquad\quad
\mathcal{D}_{\mu}F^{\alpha\mu}\,_{\nu}=0,
\qquad\quad
F^{\alpha}_{\mu\nu}F^{\beta\mu\nu}
=f^{\eta\gamma\alpha}f^{\beta\gamma\eta}.
\end{eqnarray}
These solutions are important from the physical point of view that although the plain equations in (\ref{n19}) do not solve vacuum-Einstein equations (\ref{skk1})$-$(\ref{skk3}) (see appendix), they are natural solutions of the non-Abelian Weyl$-$Yang$-$Kaluza$-$Klein gravity model. Thanks to the last equation of (\ref{n19}) and normalization of structure constants $f^{\alpha\gamma\eta}f^{\beta\gamma\eta}=(N-1)\delta^{\alpha\beta}$, the well-known quadratic invariant term $\mathcal{F}^{2}_{\alpha\beta}$ is now defined as $\mathcal{F}^{2}_{\alpha\beta}=(1-N)\delta_{\alpha\beta}$.

For $N=1$ Abelian case i.e. $G=U(1)$ and $G/H=S^{1}$ which gives the usual $5$D KK ground state $M_{4}\times S^{1}$, we dramatically obtain well-established KK field equations in the case of dilaton fields $\phi(x)=1$ where $F^{\alpha}_{\mu\nu}$ reduces to $F_{\mu\nu}$, $\mathcal{D}_{\mu}$ to ${D}_{\mu}$, $\xi^{\alpha}_{i}$ to $1$ by virtue of $\xi=\partial/\partial \ell$ with $y^{1}=\ell$, $f^{\alpha\beta\gamma}$ and $R_{ijkl}$ to $0$. We reach the following equations
\begin{eqnarray}
R_{\mu\nu}=\frac{1}{2}F_{\mu\tau}F_{\nu}\,^{\tau},
\qquad\quad
D_{\mu}F^{\mu}\,_{\nu}=0,
\qquad\quad
F_{\mu\nu}F^{\mu\nu}=0.
\end{eqnarray}
In that respect, the equations (\ref{n5}) and (\ref{n6}) vanish identically, and the remaining four reduced field equations (\ref{n1})$-$(\ref{n4}) (with the current (\ref{n0}) and the invariant term (\ref{rlag1})) are precisely consistent with those obtained from $5$D Weyl$-$Yang$-$Kaluza$-$Klein model~\cite{sibelhalil}, as expected. Besides, the equation (\ref{lfd}) is reduced to be more interesting expression $f_{\lambda}=-D_{\lambda}\mathcal{F}^{2}$  i.e.  the Yang$-$Mills force density becomes Lorentz force density of the electromagnetic theory (also see~\cite{sibelhalil}), where the nonvanishing $\mathcal{F}^{2}$ is the one of the fundamental invariants of the electromagnetism.

\section{The reduction of the stress-energy-momentum tensor}

Now we obtain the components of energy-momentum tensor of our model $\hat T_{AB}$ (\ref{set}) in the $(4+N)$-dimensional entire space. The computations  are more straightforward by taking into account non-trivial Riemann tensors (\ref{riem}) and reduced form of the quadratic curvature term $\hat R^{2}_{ABCD}$ (\ref{rlag}), however, we should recall that the noncoordinate components of the full metric (\ref{metric1}) are also used throughout this section. It is useful
to decompose components of the stress-energy tensor $\hat T_{\mu\nu}$  as well as $\hat T_{ij}$ into
trace-free $\hat{T}_{\mu\nu}^{(tf)}$ and trace $\hat{T}_{\mu\nu}^{(t)}$ parts
\begin{eqnarray}\label{se1}
\hat{T}_{\mu\nu}=\hat{T}_{\mu\nu}^{(tf)}+\hat{T}_{\mu\nu}^{(t)},
\end{eqnarray}
because not only the former looks nice but also the later can be employed to find the trace of $\hat T_{AB}$ more easily. Hence, after performing some manipulations and reorganising terms, the traceless parts are found to be
\begin{eqnarray}\label{tf1}
\hat{T}^{(tf)}_{\mu\nu}&=&{T}^{(g)}_{\mu\nu}
+3(D_{i}\xi^{\alpha}_{j}D^{i}\xi^{\beta j}
+\frac{1}{8}\xi^{\alpha i}\xi^{\gamma}_{i}\xi^{\beta j}\xi^{\eta}_{j}\mathcal{F}^{2}_{\gamma\eta})
T^{(em)\alpha\beta}_{\mu\nu}
+\frac{3}{2}\xi^{\alpha i}\xi^{\beta}_{i}{T}^{(c)\alpha\beta}_{\mu\nu}
\nonumber\\&&
+3\xi^{\alpha}_{i}\xi^{\beta}_{j}D^{i}\xi^{\gamma j}{T}^{(1)\alpha\beta\gamma}_{\mu\nu}
+\frac{1}{8}(\xi^{\alpha i}\xi^{\beta}_{i}\xi^{\gamma j}\xi^{\eta}_{j}
+4\xi^{\alpha i}\xi^{\eta}_{i}\xi^{\gamma j}\xi^{\beta}_{j}){T}^{(2)\alpha\gamma\beta\eta}_{\mu\nu}\nonumber\\&&
+\xi^{\alpha i}\xi^{\beta}_{i}{T}^{(3)\alpha\beta}_{\mu\nu}.
\end{eqnarray}
The first term of equation (\ref{tf1}) is labelled as ${T}^{(g)}_{\mu\nu}$ for the fact that it exactly gives  pure gravitational stress-energy tensor of Weyl$-$Yang gauge theory in the actual $4$D  $external$ spacetime
\begin{eqnarray}
{T}^{(g)}_{\mu\nu}=R_{\mu\lambda\sigma\rho}
R_{\nu}\,^{\lambda\sigma\rho}
-\frac{1}{4}g_{\mu\nu}R_{\tau\lambda\sigma\rho}R^{\tau\lambda\sigma\rho}.
\end{eqnarray}
Another well-known expression $T^{(em)\alpha\beta}_{\mu\nu}$ is also welcome from the equation (\ref{tf1}). That is gauge invariant energy-momentum tensor of the ordinary non-Abelian Yang$-$Mills fields which can be written as
\begin{eqnarray}
{T}^{(em)\alpha\beta}_{\mu\nu}=F^{\alpha}_{\mu\lambda}F^{\beta}_{\nu}\,^{\lambda}
-\frac{1}{4}g_{\mu\nu}F^{\alpha}_{\tau\lambda}
F^{\beta\tau\lambda}.
\end{eqnarray}
The resting non-trivial terms of equation (\ref{tf1}) have more complicated structures because they come, as might be expected, from higher-order quadratic action. We respectively identify these terms as the stress-energy tensor of the $RF^{2}$-type non-minimal coupling fields
\begin{eqnarray}
{T}^{(c)\alpha\beta}_{\mu\nu}=\frac{1}{2}(F^{\beta\rho\sigma}
F^{\alpha}_{\mu}\,^{\lambda}R_{\nu\lambda\sigma\rho}
+F^{\beta\rho\sigma}
F^{\alpha}_{\nu}\,^{\lambda}R_{\mu\lambda\sigma\rho})
-\frac{1}{4}g_{\mu\nu}F^{\alpha\tau\lambda}
F^{\beta\rho\sigma}R_{\tau\lambda\sigma\rho},
\end{eqnarray}
of the cubic fields
\begin{eqnarray}
{T}^{(1)\alpha\beta\gamma}_{\mu\nu}=\frac{1}{2}(
\mathcal{F}^{3~\alpha\beta\gamma}_{\mu\nu}
+\mathcal{F}^{3~\alpha\beta\gamma}_{\nu\mu})
-\frac{1}{4}g_{\mu\nu}\mathcal{F}^{3}_{\alpha\beta\gamma},
\end{eqnarray}
of the quartic constituent fields
\begin{eqnarray}
{T}^{(2)\alpha\gamma\beta\eta}_{\mu\nu}=
\mathcal{F}_{\mu\nu}^{4~\alpha\gamma\beta\eta}
-\frac{1}{4}g_{\mu\nu}\mathcal{F}^{4}_{\alpha\gamma\beta\eta},
\end{eqnarray}
and of the $(\mathcal{D}F)^{2}$-type interactions
\begin{eqnarray}
{T}^{(3)\alpha\beta}_{\mu\nu}&=&
\mathcal{D}_{\mu}F^{\alpha}_{\sigma\rho}\mathcal{D}_{\nu}F^{\beta\sigma\rho}
-\frac{1}{4}g_{\mu\nu}\mathcal{D}_{\sigma}F^{\alpha}_{\tau\rho}
\mathcal{D}^{\sigma}F^{\beta\tau\rho}.
\end{eqnarray}
The trace part which consists of the remaining terms of $\hat{T}_{\mu\nu}$ (\ref{se1}), on the other hand, becomes
\begin{eqnarray}\label{tsee1}
\hat{T}_{\mu\nu}^{(t)}&&=-\frac{1}{4}g_{\mu\nu}R^{2}_{ijkl}
-\frac{3}{2}\xi^{\alpha}_{i}\xi^{\beta}_{j}D^{i}\xi^{\gamma j}(\mathcal{F}^{3~\alpha\beta\gamma}_{\mu\nu}
+\mathcal{F}^{3~\alpha\beta\gamma}_{\nu\mu})
-\frac{3}{2}D_{i}\xi^{\alpha}_{j}D^{i}\xi^{\beta j}F^{\alpha}_{\mu\lambda}F^{\beta}_{\nu}\,^{\lambda}
\nonumber\\
& &
+\frac{1}{4}\xi^{\alpha i}\xi^{\beta}_{i}[\frac{1}{2}\xi^{\gamma j}\xi^{\eta}_{j}
(\mathcal{F}_{\mu\nu}^{4~\alpha\gamma\beta\eta} -2\mathcal{F}_{\mu\nu}^{4~\alpha\gamma\eta\beta})
+2\mathcal{D}_{\sigma}F^{\alpha}_{\mu\rho}
\mathcal{D}^{\sigma}F^{\beta}_{\nu}\,^{\rho}\nonumber\\
& &
-3\mathcal{D}_{\mu}F^{\alpha}_{\sigma\rho}\mathcal{D}_{\nu}F^{\beta\sigma\rho}].
\end{eqnarray}
Another reduced component of the energy-momentum tensor  is the $\hat T_{\mu i}$ which turns out to be
\begin{eqnarray}\label{se2}
\hat{T}_{\mu i}&=&\frac{1}{2}\xi^{\alpha}_{i}R_{\mu\lambda\sigma\rho}
\mathcal{D}^{\lambda}F^{\alpha\sigma\rho}
+\frac{1}{2}\xi^{\alpha}_{j}D^{j}\xi^{\beta}_{i}F^{\beta\sigma\rho}
(\mathcal{D}_{\mu}F^{\alpha}_{\rho\sigma}
+\mathcal{D}_{\rho}F^{\alpha}_{\mu\sigma})
\nonumber\\
& &+\frac{1}{4}\xi^{\alpha}_{i}\xi^{\beta j}\xi^{\gamma}_{j}
[3F^{\beta}_{\mu}\,^{\lambda}F^{\sigma\rho}_{\gamma}
\mathcal{D}_{\sigma}F^{\alpha}_{\rho\lambda}
-F^{\alpha\sigma\lambda}F^{\beta}_{\lambda}\,^{\rho}
(\mathcal{D}_{\mu}F^{\gamma}_{\sigma\rho}
+\mathcal{D}_{\sigma}F^{\gamma}_{\mu\rho})].
\end{eqnarray}
Similarly, the $i$-$j$ component of the equation (\ref{set}) can be split into a trace-free and a trace part as follows
\begin{eqnarray}
\hat{T}_{ij}=\hat{T}_{ij}^{(tf)}+\hat{T}_{ij}^{(t)},
\end{eqnarray}
where
\begin{eqnarray}\label{se3}
\hat{T}_{ij}^{(tf)}&=&
\frac{N}{4}{T}^{(g)}_{ij}
+\frac{N}{16}[\frac{1}{2}\xi^{\gamma k}\xi^{\eta}_{k}(\mathcal{F}^{4}_{\alpha\gamma\beta\eta}
+4\mathcal{F}_{\alpha\gamma\eta\beta}^{4})
+4\mathcal{D}_{\mu}F^{\alpha}_{\nu\lambda}
\mathcal{D}^{\mu}F^{\beta\nu\lambda}]{T}^{(1)\alpha\beta}_{ij}\nonumber\\
& &
+\frac{3N}{4}(\mathcal{F}^{2}_{\alpha\beta}{T}^{(2)\alpha\beta}_{ij}
+\mathcal{F}^{3}_{\alpha\beta\gamma}{T}^{(3)\alpha\beta\gamma}_{ij}).
\end{eqnarray}
The ${T}^{(g)}_{ij}$ appears formally to be $4$D stress-energy tensor of Weyl$-$Yang theory but in this case it is for the $internal$ space. That is, ${T}^{(g)}_{ij}$ is written in terms of the Riemann tensors of $internal$ space in the $N$-dimensions
\begin{eqnarray}\label{nnn1}
{T}^{(g)}_{ij}= R_{inkl}R_{j}\,^{nkl}-\frac{1}{N}g_{ij}R_{mnkl}R^{mnkl}.
\end{eqnarray}
Other unphysical energy-momentum tensors correspondingly are
\begin{eqnarray}
{T}^{(1)\alpha\beta}_{ij}&=&\xi^{\alpha}_{i}\xi^{\beta}_{j}\label{nn1}
-\frac{1}{N}g_{ij}\mathcal\xi^{\alpha}_{k}\xi^{\beta k},\\
{T}^{(2)\alpha\beta}_{ij}&=&D_{n}\xi^{\alpha}_{i}D^{n}\xi^{\beta}_{j}\label{n2}
-\frac{1}{N}g_{ij}D_{n}\xi^{\alpha}_{m}D^{n}\xi^{\beta m},\\
{T}^{(3)\alpha\beta\gamma}_{ij}&=&\frac{1}{2}\xi^{\alpha}_{n}\label{n3}
(\xi^{\beta}_{i}D^{n}\xi^{\gamma }_{j}
+\xi^{\beta}_{j}D^{n}\xi^{\gamma }_{i})
-\frac{1}{N}g_{ij}\xi^{\alpha}_{n}\xi^{\beta}_{m}D^{n}\xi^{\gamma m},
\end{eqnarray}
together with more complicated trace part
\begin{eqnarray}\label{tsee2}
\hat{T}_{ij}^{(t)}&=&
 -\frac{1}{4}g_{ij}[R^{2}_{\mu\nu\lambda\rho}
 -\frac{3}{4}\xi^{\alpha k}\xi^{\beta}_{k}( 2R_{\mu\nu\lambda\rho}F^{\alpha\mu\nu}F^{\beta\lambda\rho}
-\frac{1}{2}\xi^{\gamma l}\xi^{\eta}_{l}
\mathcal{F}^{2}_{\alpha\gamma}\mathcal{F}^{2}_{\beta\eta})]
\nonumber\\
& &-(\frac{N-4}{4})R_{inkl}R_{j}\,^{nkl}
-\frac{1}{16}\xi^{\alpha}_{i}\xi^{\beta}_{j}\big\{\frac{1}{2}\xi^{\gamma k}\xi^{\eta}_{k}
[(N+4)\mathcal{F}^{4}_{\alpha\gamma\beta\eta}
\nonumber\\
& &
+2(N-2)\mathcal{F}^{4}_{\alpha\gamma\eta\beta}]
+4(N-1)\mathcal{D}_{\mu}F^{\alpha}_{\nu\lambda}
\mathcal{D}^{\mu}F^{\beta\nu\lambda}\big\}\nonumber\\
& &
-(\frac{3N-6}{4})[D_{n}\xi^{\alpha}_{i}D^{n}\xi^{\beta}_{j}
\mathcal{F}^{2}_{\alpha\beta}
+\frac{1}{2}\xi^{\alpha}_{n}(\xi^{\beta}_{i}D^{n}\xi^{\gamma }_{j}
+\xi^{\beta}_{j}D^{n}\xi^{\gamma }_{i})\mathcal{F}^{3}_{\alpha\beta\gamma}].
\end{eqnarray}
In $5$D KK theories, $\hat T_{\mu5}$ may be interpreted as the current density with the help of conservation law $\hat D_{\mu}\hat T^{\mu}\,_{5}=0$. In our approach, it is not only easy to find  physical roles or consequences of components $\hat T_{\mu i}$ and $\hat T_{ij}$ but also to obtain analogues of those in the $4$D Einstein's theory of gravity.

Now, we can calculate the trace of the full $\hat T_{AB}$ (i.e. $\hat T$) by employing
\begin{eqnarray}\label{trace}
\hat T\equiv\hat g^{AB}\hat T_{AB}
 =\hat g^{\mu\nu}\hat T^{(t)}_{\mu\nu}+\hat g^{ij}\hat T^{(t)}_{ij}.
\end{eqnarray}
together with the equations (\ref{tsee1}) and (\ref{tsee2}). The resulting form will not be considered further, however that can be checked on due to the fact that $\hat T=-(N/4)\hat R^{2}_{ABCD}$ (from equation (\ref{set})) with equation (\ref{rlag}). It is no difficult to conjecture that, in the absence of non-Abelian gauge fields $A_{\mu}^{\alpha}=0$ we only get pure gravitational stress-energy tensor $\hat{T}_{\mu\nu}={T}^{(g)}_{\mu\nu}$ as well as $\hat{T}_{\mu i}=\hat{T}_{ij}=0$.

Attention will now be turned to investigate the  reduced energy-momentum tensors which are already obtained below for the case where the $internal$ space is the homogeneous coset space. On account of all necessary assumptions which are  mentioned and obtained in previous sections, the traceless part $\hat{T}_{\mu\nu}^{(tf)}$ in (\ref{tf1}) satisfies
\begin{eqnarray}\label{sec1}
\hat{T}_{\mu\nu}^{(tf)}&=&{T}^{(g)}_{\mu\nu}
+\frac{3}{8}(2f^{\alpha\gamma\eta}f^{\beta\gamma\eta}
+\mathcal{F}^{2}_{\alpha\beta}){T}^{(em)\alpha\beta}_{\mu\nu}
+\frac{3}{2}{T}^{(c)\alpha\alpha}_{\mu\nu}
+\frac{3}{2}f^{\alpha\beta\gamma}{T}^{(1)\alpha\beta\gamma}_{\mu\nu}
\nonumber\\
& &
+\frac{1}{8}{T}^{(2)\alpha\beta\alpha\beta}_{\mu\nu}
+\frac{1}{2}{T}^{(2)\alpha\beta\beta\alpha}_{\mu\nu}
+{T}^{(3)\alpha\alpha}_{\mu\nu},
\end{eqnarray}
and the trace part in (\ref{tsee1}) is
\begin{eqnarray}\label{sec2}
\hat{T}_{\mu\nu}^{(t)}&=&-\frac{1}{4}g_{\mu\nu}R^{2}_{ijkl}
-\frac{3}{8}f^{\alpha\gamma\eta}f^{\beta\gamma\eta}F^{\alpha}_{\mu\lambda}F^{\beta}_{\nu}\,^{\lambda}
-\frac{3}{8}f^{\alpha\beta\gamma}(\mathcal{F}^{3~\alpha\beta\gamma}_{\mu\nu}
+\mathcal{F}^{3~\alpha\beta\gamma}_{\nu\mu})
\nonumber\\
& &
+\frac{1}{2}\mathcal{D}_{\sigma}F^{\alpha}_{\mu\rho}
\mathcal{D}^{\sigma}F^{\beta}_{\nu}\,^{\rho}
-\frac{3}{4}\mathcal{D}_{\mu}F^{\alpha}_{\sigma\rho}
\mathcal{D}_{\nu}F^{\alpha\sigma\rho}
+\frac{1}{8}(\mathcal{F}_{\mu\nu}^{4~\alpha\beta\alpha\beta} -2\mathcal{F}_{\mu\nu}^{4~\alpha\beta\beta\alpha}),
\end{eqnarray}
where $R_{ijkl}^{2}=(1/8)N(N-1)$. Hence, we recognize the fact that the component $\hat{T}_{\mu\nu}$ (the summation of equation (\ref{sec1}) and equation (\ref{sec2})) is  Killing term-free equation, and the resulting expression  is more convenient to bring out the type of interactions between constituent fields. Next, the $\hat T_{\mu i}$ (\ref{se2}) yields
\begin{eqnarray}\label{tmi2}
\hat{T}_{\mu i}&=&\frac{1}{2}\xi^{\alpha}_{i}[2R_{\mu\lambda\sigma\rho}
\mathcal{D}^{\lambda}F^{\alpha\sigma\rho}
+f^{\alpha\beta\gamma}F^{\beta\sigma\rho}
(\mathcal{D}_{\mu}F^{\gamma}_{\rho\sigma}
+\mathcal{D}_{\rho}F^{\gamma}_{\mu\sigma})\nonumber\\
& &
+3F^{\beta}_{\mu}\,^{\lambda}F^{\beta\sigma\rho}
\mathcal{D}_{\sigma}F^{\alpha}_{\rho\lambda}
-F^{\alpha\sigma\lambda}F^{\beta}_{\lambda}\,^{\rho}
(\mathcal{D}_{\mu}F^{\beta}_{\sigma\rho}
+\mathcal{D}_{\sigma}F^{\beta}_{\mu\rho})].
\end{eqnarray}
Let us finally evaluate the last component $\hat T_{ij}$ of the stress-energy tensor  $\hat T_{AB}$. The equation (\ref{se3}) can be calculated as
\begin{eqnarray}\label{tij1}
\hat{T}_{ij}^{(tf)}&=&\frac{N}{16}[(\mathcal{F}^{4}_{\alpha\gamma\beta\gamma}
+4\mathcal{F}^{4}_{\alpha\gamma\gamma\beta}
+4\mathcal{D}_{\mu}F^{\alpha}_{\nu\lambda}
\mathcal{D}^{\mu}F^{\beta\nu\lambda}\nonumber\\
& &
+3\mathcal{F}^{2}_{\eta\zeta}f^{\eta\gamma\alpha}f^{\zeta\gamma\beta}
+3\mathcal{F}^{3}_{\eta\alpha\gamma}f^{\gamma\eta\beta}){T}^{\alpha\beta}_{ij}
+3\mathcal{F}^{3}_{\eta\alpha\gamma}f^{\gamma\eta\beta}{T}^{\beta\alpha}_{ij}].
\end{eqnarray}
Here, we write to show that
\begin{eqnarray}
{T}^{\alpha\beta}_{ij}&=&\xi^{\alpha}_{i}\xi^{\beta}_{j}\label{g1}
-\frac{1}{N}g_{ij}\delta^{\alpha\beta}.
\end{eqnarray}
From another point of view, the equation (\ref{tsee2}) which is a trace part of $\hat{T}_{ij}$ is equivalent to
\begin{eqnarray}\label{tij2}
\hat{T}_{ij}^{(t)}&=&
 -\frac{1}{4}g_{ij}[R^{2}_{\mu\nu\lambda\rho}
 -\frac{3}{2} R_{\mu\nu\lambda\rho}F^{\alpha\mu\nu}F^{\alpha\lambda\rho}
+\frac{3}{8}(\mathcal{F}^{2}_{\alpha\beta})^{2}]\nonumber\\
& &
-(\frac{3N-6}{16})
f^{\gamma\eta\beta}(\xi^{\alpha}_{i}\xi^{\beta}_{j}+\xi^{\alpha}_{j}\xi^{\beta}_{i})
\mathcal{F}^{3}_{\eta\alpha\gamma}
-\frac{1}{32}\xi^{\alpha}_{i}\xi^{\beta}_{j}
[(N+4)\mathcal{F}^{4}_{\alpha\gamma\beta\gamma}\nonumber\\
& &
+4(N-2)\mathcal{F}^{4}_{\alpha\gamma\gamma\beta}
+8(N-1)\mathcal{D}_{\mu}F^{\alpha}_{\nu\lambda}
\mathcal{D}^{\mu}F^{\beta\nu\lambda}
\nonumber\\
& &+6(N-2)f^{\eta\gamma\alpha}f^{\zeta\gamma\beta}
\mathcal{F}^{2}_{\eta\zeta}]
-(\frac{N-4}{4})R_{inkl}R_{j}\,^{nkl},
\end{eqnarray}
where $R_{inkl}R_{j}\,^{nkl}=(1/8)(N-1)g_{ij}$. The compatibility is also welcome here for the $5$D KK picture $N=1$ where $\xi^{\alpha}_{i}=1$ and $R_{ijkl}=0$. The equations (\ref{sec1}), (\ref{sec2}), (\ref{tmi2}) and (\ref{tij2}) are exactly identical to those obtained from  our previous gravity model~\cite{sibelhalil}, and the equation (\ref{tij1}) goes explicitly to zero. It should be emphasised that, we use well-known tensorial rule in (\ref{tmi2}), the inner product of a symmetric and an antisymmetric tensor vanishes. We finish with one clear remark that, the equation (\ref{trace}) is also completely true for the coset space case in view of the equations (\ref{sec2}), (\ref{tij2}) and (\ref{rlag1}).

\section{Discussion and conclusions}

In this paper, we have successfully established a non-Abelian version of the spinless and torsionless Weyl$-$Yang gravitational gauge model by considering higher-dimensional pure KK theories. In other words, we have shown that it is also possible to construct a unification between the gauge theory of gravitation and Yang$-$Mills theories in the context of the Weyl$-$Yang$-$Kaluza$-$Klein picture. The resulting formalism is mathematically attractive and naturally turns out to be a generalization and extension of our previous model~\cite{sibelhalil} which corresponds to a $5$D dimensional Weyl$-$Yang theory in the $U(1)$ Abelian case, and it seems likely to be a much more complete theory with gravitational source term and Yang$-$Mills force density than the ordinary KK approach in both Abelian and non-Abelian cases.

In this sense, the field equations of the theory are investigated  dealing with $(4+N)$-dimensional Stephenson$-$Kilmister$-$Yang quadratic Lagrangian without supplementary spin-matter fields by making use of the method of Palatini variational principle due to the fact that there is no relation between the affine connection and the metric. We have brought forward a matter and radiation Lagrangian which is carrying energy-momentum but not possessing any spin tensor leading to a source current term. By employing the generalized KK approaches and applying traditional dimensionality-reduction mechanism to the sets of field equations which are suggested by Fairchild~\cite{fairchild} as well as the quadratic invariant, we have simultaneously  obtained the desired dimensionally reduced equations which are the field equations, energy-momentum tensors and modified $4$D  Weyl$-$Yang$+$Yang$-$Mills action. As expected, these resulting equations appear to be not only  more general but also more complicated, because of the higher-order quadratic action and the third-order Yang's field equations, than the well-established field equations and energy-momentum tensors of the non-Abelian theory which come from Einstein's theory of relativity. In particular, we have analyzed a homogeneous coset space reduction (without considering  consistency problem in equation (\ref{kii}) i.e. $\Psi^{\alpha\beta}(y)=\delta^{\alpha\beta}$) of the theory which gives more physical and understandable results, to compare and discuss those obtained from our previous gravity model and classical non-Abelian theories. Hence, we have concluded that in the reduction of the sets of field equations  from $(4+N)$ to four/five dimensions, we have obviously obtained  usual $4$D/$5$D Weyl$-$Yang equations. The general structure and the well-established equations of the ordinary non-Abelian KK are given in the appendix as well.

The consequences of our analysis are that the field equations naturally contain the generalized $1$-forms gravitational source term  which has been missing in the literature so far  (one of the main problem of the Weyl$-$Yang theory) and the term which describes generalized $4$-forms density of the Yang$-$Mills force which generally emerges from the higher-dimensional geodesic equations just as the Abelian case~\cite{sibelhalil}. Besides, as a result of the alternative Yang's equation (\ref{fe2}), we have explicitly showed that the field equations can dramatically be written in terms of some particular combinations of those obtained from ordinary non-Abelian KK theory. As mentioned before, there is no appropriate vacuum-Einstein solution of that theory, since the $internal$ space has to be curved for any non-Abelian group. We have proved that the plain equations in (\ref{n19}) are not ground state solutions of  KK theories with non-Abelian gauge fields (\ref{Ricci}). However, they are natural solutions of the non-Abelian Weyl$-$Yang$-$Kaluza$-$Klein gravity model for the coset space case. On the other hand, the energy-momentum tensor includes the well-known stress-energy tensors of the Weyl$-$Yang and Yang$-$Mills theory together with various types of constituent fields. This situation supports the Fairchild's idea that the equation (\ref{set}) is interpreted as the energy-momentum tensor of the theory. It should also be emphasised that just as  the Maxwell tensor of electrodynamics, the $4$D matter-spin tensor term is induced
from  higher dimensions that matter carrying energy-
momentum but not possessing any spin tensor which guarantees that energy-momentum is covariantly conservative.  Instead of this unphysical case we may consider Kaluza's idea that the universe in higher-dimensions is fully empty which means that we can also take $\hat T_{AB}=0$ in (\ref{set}). Although this idea may present more natural field equations, it also leads to not-Riemannian solutions~\cite{fairchild}, and that case has not been discussed in this paper.

We have commented in our previous work~\cite{sibelhalil} that ``The Stephenson$-$Kilmister$-$Yang-type of a $5$D model (the $5$D Weyl$-$Yang theory) with the KK ansatz and through a KK-type of dimensional reduction mechanism  appears to be more complete compared to that of the standard KK model by accommodating terms that can be identified as the Lorentz force density''. This conclusion is actually based on the fact that the gauge theories of gravitation have more physical phenomena than Einstein's theory of gravity and ``the third order equation is more natural than the second order one''~\cite{yang}. Our model has effectively supposed that idea with the generalized gravitational source term which does not exist in the literature so far and Yang$-$Mills force which is exactly equivalent to the negative gradient of a Yang$-$Mills quadratic invariant. As a final remark, we can further consider the theory with nonvanishing massless scalar fields $\phi (x)$~\cite{Awada,Appequist} which play an important role in supergravity and many other theories to understand all the physical implications of our approach. We can also investigate the field equations by varying appropriate form of the action (\ref{rlag1}) with respect to the variables. We shall discuss these constructions in detail in another study.

\section*{Acknowledgments}
I would like to thank S. Ba\c{s}kal  for bringing  references~\cite{oktem} and~\cite{dereli} into my attention.
\section*{Appendix: the non-Abelian Kaluza$-$Klein theory in the anholonomic frame}

A brief review to the major steps of non-Abelian KK framework unifying gravitation and Yang$-$Mills theories in more than five dimensions is appropriate. First, let us remember some basic notions of that theory in the usual way for convenient reading. In what follows, the Greek indices $\mu,\nu,...=0,...,3$ refer to the $external$ $4$D flat Minkowski (Ricci flat) spacetime $M_{4}$, admitting  the metric ${g}_{\mu \nu}(x)$ with usual signature  and  collectively coordinates $x\in M_{4}$. The Latin indices $i,j,...=5,...,(4+N)$ refer to the curved $internal$ $N$-dimensional  compact sub-space $M_{N}$ (such as simply the hypersphere or hypertorus), admitting the metric ${g}_{ij}(y)$ with Euclidean signature and collectively  coordinates $y\in M_{N}$, whereas the Latin capital indices $A,B,...=0,...,3,5,...,(4+N)$ refer to
the whole $(4+N)$-dimensional Minkowski space $M_{4+N}$, with the metric $\hat g_{AB}(x,y)$ and associated with the collectively event $z\in M_{4+N}$, $z=(x,y)$. The quantities with/without the hat symbol demonstrate the ones in the $(4+N)$-dimensional entire space/in the usual $4$D $external$ space. The stable ground state of the generalized 5D KK theory is assumed to be, at least locally, a direct space product of the form $M_{4+N}=M_{4}\times M_{N}$ which is a trivial principal bundle over a $M_{4}$ with fibres $M_{N}$, instead of assuming to be only $M_{4+N}$. Finally, the Greek indices $\alpha,\beta,...$ refer to any compact isometry (Lie) group $G$ of $M_{N}$, running over the rank of $G$, i.e. $\alpha,\beta,...=1,...,dim(G)$. It does not matter whether the indices of internal space are in upper or lower form. The isometries of the $internal$ manifold produce linearly independent space-like Killing vector fields $\xi^{i}_{\alpha}(y)$ each corresponding to a metric symmetry in an elegant way. The symmetries of $M_{N}$  appear to be gauge group in the real 4D world for the effective observer as to be the 5D KK approach. The Killing vector fields $\xi^{i}_{\alpha}(y)$ respectively satisfy the Lie's equation
\begin{eqnarray}\label{ke1}
[\xi_{\alpha},\xi_{\beta}]^{i}\equiv\xi^{j}_{\alpha}\partial_{j}\xi^{i}_{\beta}
-\xi^{j}_{\beta}\partial_{j}\xi^{i}_{\alpha}&=&-f_{\alpha\beta} \,^{\gamma}\xi^{i}_{\gamma},
 \end{eqnarray}
corresponding to the Lie algebra by the Lie bracket and the following isometry condition
\begin{eqnarray}\label{ke2}
\mathcal{L}_{\xi}g_{ij}\equiv\xi^{\alpha k}\partial_{k}g_{ij}+g_{ik}\partial_{j}\xi^{\alpha k}+g_{jk}\partial_{i}\xi^{\alpha k}=0,
\end{eqnarray}
which gives Killing's equation $D_{(i}\xi^{\alpha}_{j)}=0$. Here, $f_{\alpha\beta} \,^{\gamma}$ are the real antisymmetric $f_{\alpha\beta} \,^{\gamma}=-f_{\beta\alpha} \,^{\gamma}$ structure constants of $G$. The $(4+N)$-dimensional metric $\hat g_{AB}(x,y)$ can be written in terms of $1$-forms massless gauge fields (Yang$-$Mills vector bosons) $A^{\alpha}_{\mu}(x)$ of the arbitrary gauge group $G$ and the Killing vector fields $\xi^{i}_{\alpha}(y)$ in the higher-dimensional spacetime $M_{4}\times M_{N}$ as follows
\begin{equation}\label{metric}
\hat g_{AB}(x,y)=\left( \begin{array}{c|c}
    {g}_{\mu \nu}(x)+ {g}_{ij}(y)\xi^{\alpha i}(y)\xi^{\beta j}(y)A^{\alpha}_{\mu}(x)A^{\beta}_{\nu}(x)  &   {g}_{ij}(y)\xi^{\alpha i}(y)A^{\alpha}_{\mu}(x) \\
    \hline
     {g}_{ij}(y)\xi^{\beta j}(y)A^{\beta}_{\nu}(x)    &  {g}_{ij}(y) \\
   \end{array} \right)
\end{equation}
It is very useful and convenient to make the metric $\hat g_{AB}(x,y)$ block diagonal for calculations. It can be achieved by choosing the basis in the so-called anholonomic (noncoordinate, horizontal lift) basis~\cite{Cho} with
\begin{eqnarray}
\hat E^{\mu}(x)=dx^{\mu},\qquad\qquad
\hat E^{i}(x,y)=dy^{i}+\xi^{\alpha i}(y)A^{\alpha}_{\mu}(x)dx^{\mu}.
\end{eqnarray}
The dual basis can be found by the help of the identity $\hat E^{A}\hat \iota_{B}=\hat\delta^{A}\,_{B}$ in the following forms
\begin{eqnarray}\label{dual}
\hat \iota_{\mu}(x,y)=\partial_{\mu}-\xi^{\alpha i}(y)A^{\alpha}_{\mu}(x)\partial_{i}, \qquad\qquad
\hat \iota_{i}(y)=\partial_{i},
\end{eqnarray}
where we use $\partial/\partial x^{\mu}=\partial_{\mu}$ and $\partial/\partial y^{i}=\partial_{i}$ shortly. Hence, the metric components in equation (\ref{metric}) reduce to a simple form in which both ${g}_{\mu \nu}$ and ${g}_{ij}$ are now diagonal so that the $(4+N)$-dimensional metric $\hat g_{AB}(x,y)$ becomes
\begin{eqnarray}\label{metric1}
\hat g_{AB}(x,y)=\left( \begin{array}{c|c}
    {g}_{\mu \nu}(x) &  0  \\
      \hline   0    &  {g}_{ij}(y) \\
   \end{array} \right).
\end{eqnarray}
Now, it is very easy to raise and lower indices by taking into account the form of $\hat g_{AB}(x,y)$ in equation (\ref{metric1}). By employing necessary relations that can be found in~\cite{misner} for this basis, the nonvanishing components of the $(4+N)$-dimensional curvature tensor $\hat{R}^{A}\,_{BCD}$ decompose into
\begin{eqnarray}\label{riem}
&&\hat{R}^{\mu}\,_{\nu\lambda\sigma}={R}^{\mu}\,_{\nu\lambda\sigma}
-\frac{1}{4}{g}_{ij}\xi^{\alpha i}\xi^{\beta j}
(2F^{\alpha\mu}\,_{\nu}F^{\beta}_{\lambda\sigma}
+F^{\alpha\mu}\,_{\lambda}F^{\beta}_{\nu \sigma}
-F^{\alpha\mu}\,_{\sigma}F^{\beta}_{\nu \lambda}), \nonumber\\
&&\hat{R}^{i}\,_{\nu\lambda\sigma}=\frac{1}{2}\xi^{\alpha i} \mathcal{D}_{\nu}F^{\alpha}_{\lambda\sigma},\nonumber\\
&&\hat{R}^{i}\,_{\nu j\sigma}=\frac{1}{2}D_{j}\xi^{\alpha i} F^{\alpha}_{\nu\sigma}
-\frac{1}{4}{g}_{ij}\xi^{\alpha i}\xi^{\beta j}F^{\alpha}_{\sigma\tau}F^{\beta\tau}\,_{\nu},\\
&&\hat{R}^{\mu}\,_{\nu ij}=D_{j}\xi^{\alpha }_{i} F^{\alpha\mu}\,_{\nu}
+\frac{1}{4}\xi^{\alpha}_{i}\xi^{\beta}_{j}(F^{\alpha\mu\tau}F^{\beta}_{\tau\nu}
-F^{\beta\mu\tau}F^{\alpha}_{\tau\nu}),\nonumber\\
&&\hat{R}^{i}\,_{jkl}={R}^{i}\,_{jkl}.\nonumber
\end{eqnarray}
We present these expressions, since they are the basis of what follows. For completeness,  the electromagnetic field strength tensor $F^{\alpha}_{\mu\nu}(x)$ and the Yang-Mills total covariant derivative in equation (\ref{riem}) are explicitly given as
\begin{eqnarray}
F^{\alpha}_{\mu\nu}=\partial_{\mu}A^{\alpha}_{\nu}-\partial_{\nu}A^{\alpha}_{\nu}\label{fs}
+f_{\beta\gamma}\,^{\alpha}A^{\beta}_{\mu}A^{\gamma}_{\nu},\qquad
\mathcal{D}_{\mu} F^{\alpha}_{\nu\lambda}=D_{\mu} F^{\alpha}_{\nu\lambda}
+f_{\beta\gamma}\,^{\alpha}A^{\beta}_{\mu}F^{\gamma}_{\nu\lambda}.
\end{eqnarray}
The reduced forms of Ricci tensor $\hat{R}_{AB}$ are, on the other hand, expressed from the equation (\ref{riem}) as
\begin{eqnarray}\label{Ricci}
\hat{R}_{\mu\nu}&\equiv&\mathcal{P}_{\mu\nu}= R_{\mu\nu}-\frac{1}{2}{g}_{ij}\xi^{\alpha i}\xi^{\beta j}F^{\alpha}_{\mu\lambda}F^{\beta}_{\nu}\,^{\lambda},\label{skk1}\\
\hat{R}_{i\nu}&\equiv&\mathcal{Q}_{i\nu}=\xi^{\alpha}_{i} \mathcal{D}_{\mu}F^{\alpha\mu}\,_{\nu},\label{skk2}\\
\hat{R}_{ij}&\equiv&\mathcal{U}_{ij}= R_{ij}+\frac{1}{4}\xi^{\alpha}_{i}\xi^{\beta}_{j}F^{\alpha}_{\lambda\tau}F^{\beta\lambda\tau}\label{skk3}.
\end{eqnarray}
Here, $R_{\mu\nu}$ and $R_{ij}$ are the  $4$D  Ricci tensors of $external$ and $internal$ spaces, respectively. The Ricci tensors contracting to get Ricci scalar, we find the curvature invariant corresponding to the non-Abelian ansatz (\ref{metric}) is given as
\begin{equation}\label{scalar}
\hat R(x,y) = R(x)+R(y)-\frac{1}{4}{g}_{ij}(y)\xi^{\alpha i}(y)\xi^{\beta j}(y)F^{\alpha}_{\lambda\tau}(x)F^{\beta\lambda\tau}(x),
\end{equation}
where $R(x)$ and $R(y)$ are scalar curvatures in four  and  $N$ dimensions, respectively~\cite{straumann}. As it is known, to obtain conventional form of the $4$D gauge fields in equation (\ref{scalar}), i.e. to construct a desired $4$D field theory which only includes the graviton and massless Yang-Mills fields, we must select and normalize the Killing vector fields such as
\begin{equation}\label{kill}
{g}_{ij}(y)\xi^{\alpha i}(y)\xi^{\beta j}(y)=c\delta^{\alpha\beta},
\end{equation}
for some constant $c$. We also add an appropriate cosmological constant term $\hat\Lambda$ to the action of the theory to avoid contributions from non-zero $R(y)$ term. Then the equation (\ref{scalar}) is completely independent of the $internal$ coordinates. Unfortunately, the equation (\ref{kill}) is generally not true for any choice of $internal$ space, and we need to write a matrix term $\Psi^{\alpha\beta}(y)$ to the right-hand side of the equation (\ref{kill}) rather than $\delta^{\alpha\beta}$
\begin{equation}\label{kii}
\xi^{\alpha i}(y)\xi^{\beta}_{i}(y)=\Psi^{\alpha\beta}(y),
\end{equation}
which however leads to a well-known consistency problem in the non-Abelian KK framework. Thanks to supergravity theories, one can prove that $\Psi^{\alpha\beta}(y)=\delta^{\alpha\beta}$ by making use of supergravities~\cite{Duff84,Duff85}.

Now, we can obtain an appropriate vacuum solution looking for the equations of motions $\hat{R}_{AB}=0$ just as classical 5D KK theory. The last equation (\ref{skk3}) does not vanish $\hat{R}_{ij}\neq 0$ (not Ricci flat) because the $internal$  space has to be curved for any non-Abelian group. However, by adding suitable matter fields to the Hilbert$-$Einstein Lagrangian, we can obtain an acceptable vacuum solution of the form $M_{4}\times M_{N}$ which is called $spontaneous~compactification$ by Cremmer and Scherk~\cite{Cremmer,Cremmer1,Cremmer2}.

\end{document}